\documentclass{elsart}

\usepackage{amssymb}
\usepackage{graphics}
\usepackage{graphicx}
\usepackage{graphicx,colordvi}
\usepackage[dvips]{color}

\begin{document}

\begin{flushright}
  TRI-PP-04-19
\end{flushright}

\begin{frontmatter}



\title{Effect of kinematics on final state interactions  in
       ($e,e'p$) reactions}

\author[TRIUMF]{C.~Barbieri}\ead{barbieri@triumf.ca},
\author[Basel]{D.~Rohe},
\author[Basel]{I.~Sick}~and~
\author[NIKHEF]{L.~Lapik\'{a}s}
\address[TRIUMF]{TRIUMF, 4004 Wesbrook Mall, Vancouver, 
          British Columbia, Canada V6T 2A3 }
\address[Basel]{Dept. f\"ur Physik und Astronomie, Universit\"at Basel, CH-4056, Switzerland}
\address[NIKHEF]{NIKHEF, P.O. Box 41882, 1009 DB Amsterdam, The Netherlands}

\begin{abstract}
Recent data from experiment E97-006 at TJNAF using the ${}^{12}{\rm C}(e,e'p)$
reaction at very large missing energies and momenta are compared to
a calculation of two-step rescattering.
 A comparison between parallel and perpendicular kinematics suggests
that the effects of final state interactions can be strongly reduced in the
former case.
\end{abstract}

\begin{keyword}
electron scattering \sep short range correlations

\PACS 25.30.Fj \sep 25.30.Dh \sep 21.60.-n \sep 21.10.Pc \sep 21.10.Jx.
\end{keyword}
\end{frontmatter}


 Short-range and tensor correlations (SRC) have long been known to be one
of the major elements influencing the dynamics
of nuclear systems~\cite{PandSick97,DB04}.
 Their main effects consist in shifting a sizable amount of spectral
strength,  10-20\%~\cite{bv91}, to very high energies
and momenta.
This results in an equally large depletion of quasi-particle
orbitals~\cite{Marcel},
which is fairly independent of the given shell and the nuclear mass number.
 Several theoretical studies,
based on realistic phenomenological NN interactions,
have suggested that most of the correlated  strength is found along
a ridge in the momentum-energy plane ($k$-$E$) which spans a region of  
several hundred MeV/c (MeV)~\cite{Ciofi90,Benhar,WHA} 
which corresponds to large missing momenta ($p_m$) and energies ($E_m$)
in knock out cross sections.
This contribution to the spectral function is also responsible
for most of the binding energy of nuclear systems~\cite{Wim03}.
The main characteristics predicted by these calculations are consistent
with recent experimental data~\cite{danielaPRL,Frick04}, which will be
considered further below.

An accurate  experimental determination of the correlated strength
in the hole spectral function $S^h(p_m,E_m)$ would represent an important
advance in our understanding of SRC, but locating this strength
at both large $E_m$ and $p_m$ is difficult.
 Early attempts done by means of ($e,e'p$) reactions found an enormous
background generated by final state interactions (FSI), see for example
Refs.~\cite{MIT-BATES1,MIT-BATES2}.
 The principal reason is that the  correlated strength is spread over an
energy range of several hundred MeV, so the total density of the
spectral function is very low. 
 In this energy regime multi-nucleon processes, beyond the direct knock out,
are possible~\cite{Takaki89} and can induce large shifts in the
missing energies and momenta, moving strength to regions where the direct signal
is much smaller and therefore submerges it.
 Calculations of FSI effects were done by several
groups~\cite{Takaki89,VivianPV,NicolaevGL,Ciofi4He,RyckebuschO16}.
 In general, theory predicts larger effects when the transverse
structure functions that enter the expression of the ($e,e'p$)
cross section dominate the longitudinal one.
 Interference effects between FSI
and initial state correlations (IC) can also play a role~\cite{Ciofi4He}.
 The results of Refs.~\cite{VivianPV,Ryckebusch03} suggests that multiple
rescattering contributions (more than two-steps) are relatively small
in light nuclei and when parallel kinematics are considered%
\footnote{In this work we refer to `parallel' and `perpendicular' kinematics
in terms of the angle between the momentum transfered by the electron ${\bf q}$
and the momentum of the initial nucleon ${\bf p}_i=-{\bf p}_m$
(as opposed to the final proton ${\bf p}_f$).
 This definition is more restrictive in the limit of high momentum transfer,
where ${\bf q}$ and ${\bf p}_f$ tend to be collinear.}.
However, it is clear that the identification of the correlated strength
cannot be achieved unless the contributions from FSI can be reduced,
by the choice of kinematics, to a size where they can be handled
by calculations.
This issue has been recently discussed in detail in Ref.\cite{IngoElba}.
It was already 
addressed in  the proposal of experiment E97-006 \cite{E97proposal} 
where a Monte Carlo simulation and kinematical arguments lead
to the suggestion that the best chance for an identification of SRC
occurs in parallel kinematics.
The latter tend to be less sensitive to meson exchange currents
(MEC) -- which involve transverse excitations -- and are more clean due
to the high momentum that is required for the detected proton.
 New data were subsequently taken by the \hbox{E97-006} collaboration
at Jefferson Lab~\cite{danielaDiss,danielaPRL,Frick04} for
a set of nuclei ranging from carbon to gold. 
Both optimal (parallel) and perpendicular kinematics were used,
to provide useful data for investigating the reaction mechanism.

In the energy regime of the JLab experiment,
the relevant contribution to FSI is identified
with two-step rescattering. This has been studied recently
in Ref.~\cite{BaLap04} using a semiclassical model.
The particular approach taken there already has the advantage
of describing the distortion due to FSI in terms of the full
one hole spectral function.
 This takes into account both the momentum and energy distribution 
of the original correlated strength,
which is of importance for the proper description
of the response~\cite{SpectFunctPRL}.
In this letter we carry out a first comparison of the prediction of 
Ref.~\cite{BaLap04} with the data of Ref.~\cite{danielaDiss} for
the nucleus ${}^{12}{\rm C}$.


We consider contributions to the experimental yield that come from two-step
mechanisms in which the knock out reaction $(e,e'a)$ for a nucleon $a$ is
followed by a scattering process from a nucleon in the medium, $N'(a,p)N''$,
eventually leading to the emission of the detected proton and the undetected
nucleon $N''$.
 Three channels are considered in the present work, in which $a$
represents either a proton (with $N'=p$ or $n$) or a neutron (with $N'=p$).
The semi-exclusive cross section for these events was calculated
according to Ref.~\cite{BaLap04} as
\begin{eqnarray}
  \lefteqn{
    { d^6  \sigma_{rescatt.}
     \over
     dE_0 \; d\Omega_{\hat{k}_o}  dE_f \; d \Omega_{\hat{p}_f} } 
 ~=~  \sum_{a , N' = 1 , 2 , 3}
    \int d {\bf r}_1 \int d {\bf r}_2 \int_{0}^{\omega} d T_a
    }
 \hspace{.4cm}    & &
\nonumber  \\
 &\;& \times 
  \rho_N({\bf r}_1) \; 
    { K \; S^h_a(|{\bf q}-{\bf p}_a|,\omega-E_a) \; \sigma^{cc}_{ea}
    \over
    M \; ( {\bf r}_1 - {\bf r}_2 )^2 } 
    g_{aN'}(|{\bf r}_1 - {\bf r}_2|) \; \; \;
\label{eq:TotRes}
 \\
 &\;& \; \times 
 P_T(p_a; {\bf r}_1 , {\bf r}_2 )
\rho_{N'}({\bf r}_2) \; 
    { d^3  \sigma_{a N'}
    \over
    dE_f \; d \Omega_{\hat{p}_f} } \;
P_T(p_f; {\bf r}_2 , \infty) \; ,
\nonumber 
\end{eqnarray}
where $(E_o,{\bf k}_o)$,  $(E_f,{\bf p}_f)$ and $(\omega,{\bf q})$
represent the four-momenta of the detected electron, the final proton
and the virtual photon, respectively.
Eq.~(\ref{eq:TotRes}) assumes that the intermediate particle $a$ is
generated in plane wave impulse approximation (PWIA)
by the electromagnetic current at a point ${\bf r}_1$
inside the nucleus. Here $K=|{\bf p}_a|E_a$ is a phase space factor,
$S^h_a(k,E)/M$ is
the spectral function of the hit particle $a$ normalized to one [i.e.,
$M=N$($Z$) if $a$ is a neutron (proton)]
and $\sigma^{cc}_{ea}$ the electron-nucleon cross section.
 In the calculations below we used the $\sigma^{cc}$ prescription
discussed in Ref.~\cite{danielaDiss}, which is obtained by employing
the on-shell current also for off-shell protons%
\footnote{Preferably one uses a prescription to extrapolate
the on-shell cross section to the off-shell case while preserving energy 
and current conservation.
 The analysis of  several possible prescriptions carried out
in Ref.~\cite{danielaDiss} found that the best agreement between the data
of different kinematics is obtained by avoiding any of the {\em ad hoc}
modifications of the kinematics at the electromagnetic vertex
as suggested in\ Ref.~\cite{deForest}.}.
%
The pair distribution functions
$g_{aN'}(|{\bf r}_1 - {\bf r}_2|)$ account for the joint probability of
finding a nucleon N' at  ${\bf r}_2$ after the particle $a$ has been struck
at ${\bf r}_1$~\cite{gNN}.
The integration over the kinetic energy $T_a$ of the intermediate nucleon $a$
ranges from 0 to the energy $\omega$ transfered by the electron.
The point nucleon densities $\rho_N({\bf r})$ were derived from experimental 
charge distributions by unfolding the proton size~\cite{density_tables}
and
the factor $P_T(p; {\bf r}_1 , {\bf r}_2)$ gives the transmission
probability that the struck particle $a$ propagates, without any
interactions,
to a second point ${\bf r}_2$, where it scatters from the nucleon $N'$ with
cross section $d^3  \sigma_{a N'}$.
The latter was obtained by adding the constraints of Pauli blocking
to the vacuum nucleon-nucleon (NN) cross section and accounting for
the Fermi distribution of the hit nucleon~\cite{BaLap04}.
 This approach is accurate for the kinematics of this work since
at large nucleon momenta the dispersion effects of the medium
become negligible.
%


\begin{figure}[!t]
\vspace{.2in}
  \begin{center}
    \includegraphics[width=0.7\linewidth]
                  {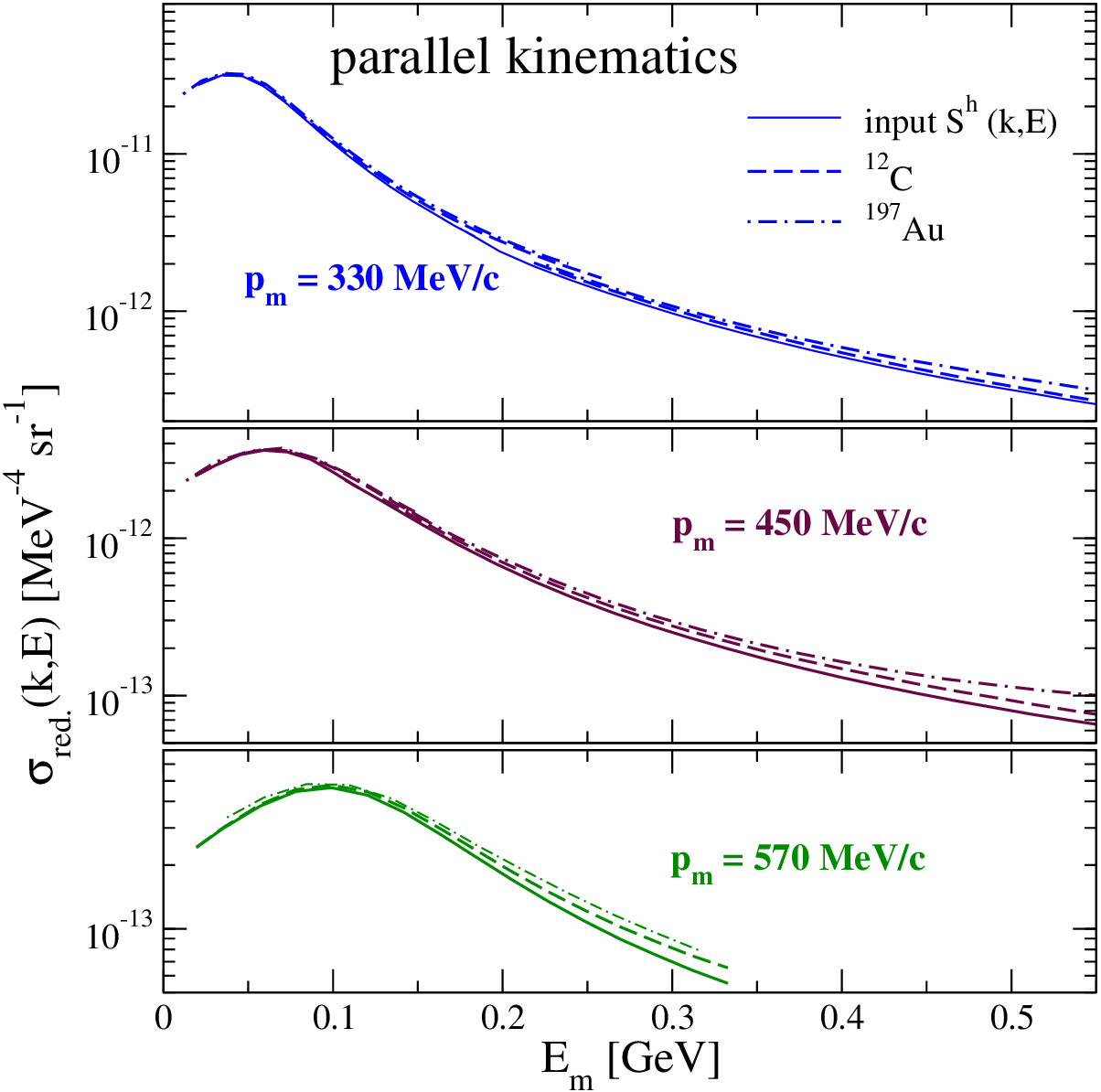}
    \caption{ \label{fig:parall_teo}
       (Color online) 
      Theoretical reduced cross section in the correlated region
     obtained in parallel kinematics for ${}^{12}{\rm C}$ (dashed-line)
     and ${}^{197}{\rm Au}$ (dot-dashed line) targets.
     The results for gold have been normalized to the number of 
     protons in carbon, for comparison.
     The full line shows the model spectral function, Eq.~(\ref{eq:Shtotal}).
     }
  \end{center}
\end{figure}

 Figure~\ref{fig:parall_teo} shows the reduced cross section
for both ${}^{12}{\rm C}$  and ${}^{197}{\rm Au}$ targets defined 
as
$\sigma_{red}(p_m,E_m)=(\sigma_{PWIA}+\sigma_{rescatt.})/(|p_fE_f|\sigma^{cc}_{ep}T)$,
where $\sigma_{PWIA}$ is the PWIA cross section of the direct process
and $T$ the nuclear transparency.
For the case of gold, the results have been normalized according to
the number of protons in ${}^{12}{\rm C}$, for comparison.
%
Eq.~(\ref{eq:TotRes}) predicts small contributions of FSI for
parallel kinematics, with a slight increase at very large
missing energies~\cite{BaLap04}.
%
%
It is important to observe that the main reason for the small effects
of rescattering  obtained in this calculation is kinematical in origin.
Due to rescattering events, the emitted nucleon looses part of
its initial energy by knocking out a second particle.
 The requirement of small angles between the momenta ${\bf q}$ and
${\bf p}_i$ implies even larger energies $T_a$ (i.e. small $E_m$)
and missing momenta for the intermediate nucleon.
Therefore, the rescattered strength is sampled in regions where the
correlated spectral function is smaller than for the direct process.
 For analogous reasons, multiple rescattering effects become even less
important, as seen in Ref.~\cite{VivianPV} for perfectly parallel kinematics.

 Recently, the experimental strength measured for ${}^{12}{\rm C}$ in
parallel kinematics was compared to theory~\cite{danielaPRL,Frick04}.
For missing energies up to 200~MeV, the summed 
strength measured turned out to agree with theoretical predictions.
 Also, the ridge-like shape of strength distribution was observed
except that the position of the peak was found at lower missing energies
than predicted by theory.
This gives confidence that, for the first time, effects of the high momentum
components attributed to SRC could be observed without being overwhelmed
by the distortion of FSI.
 However, a quantitative understanding of the reaction mechanism
is still needed.

\begin{figure}[!t]
\vspace{.15in}
  \begin{center}
    \includegraphics[width=0.65\linewidth]
                                  {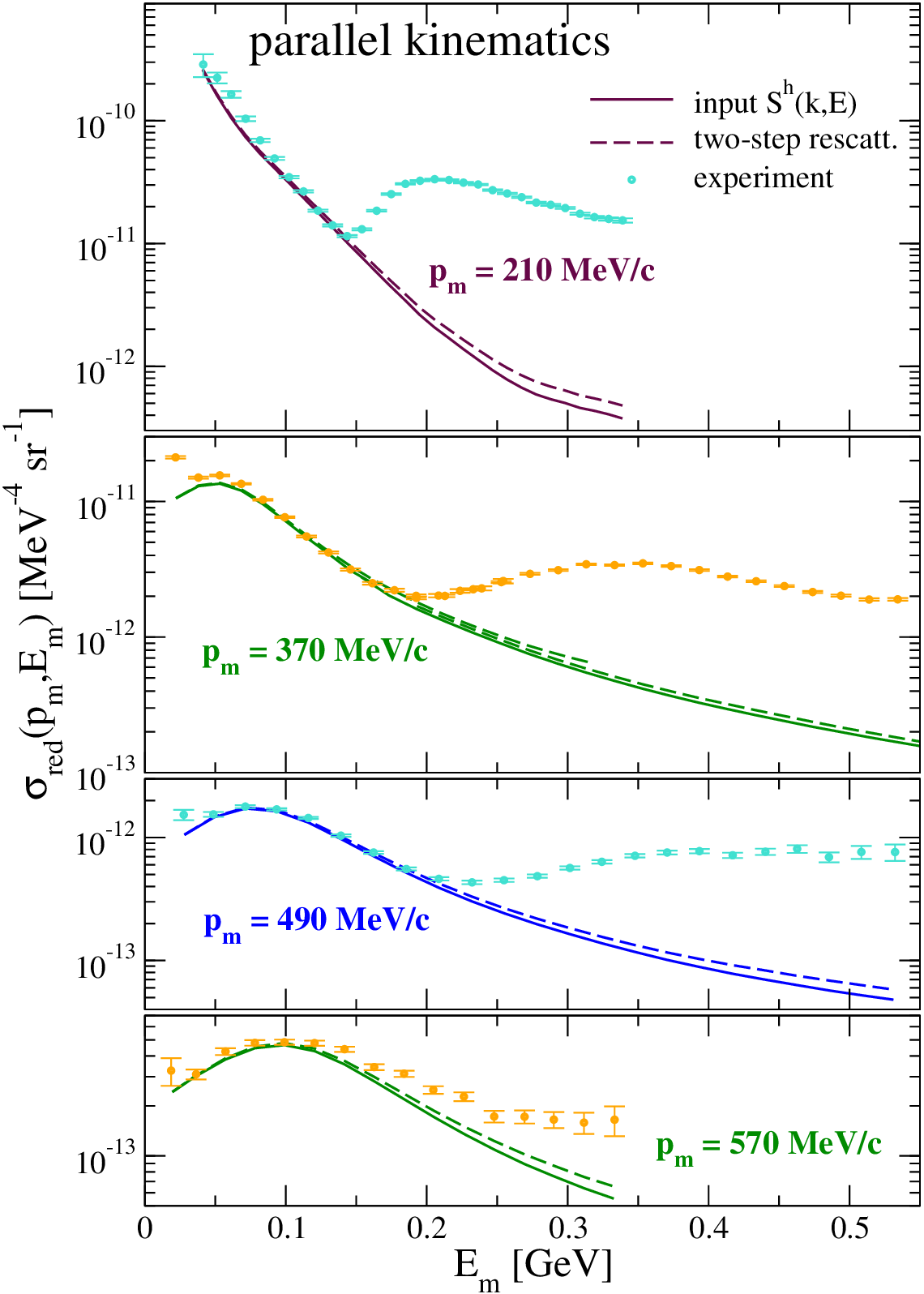}
    \caption{ \label{fig:parall_exp}
       Theoretical results for the reduced cross section of
       ${}^{12}{\rm C}$  obtained in
      parallel kinematics (dashed line) compared to the experimental
      results of Ref.~\cite{danielaDiss}.
       The full line shows the model spectral function of
      Eq.~(\ref{eq:Shtotal}) employed in the calculations.
}
\end{center}
\end{figure}

In order to make a meaningful comparison between the experiment and
the theoretical predictions for rescattering we need a proper ansatz
for the undistorted spectral function, $S^h(k,E)$ in Eq.~(\ref{eq:TotRes}).
At low energies and momenta we employed the correlated part
of the spectral function 
$S^h_{LDA}(k,E)$ in Ref.~\cite{Benhar} which was 
obtained using local density approximation (LDA), and combined it with a proper
scaled spectral function $S{^h}_{WS}(k,E)$ derived from a Wood--Saxon potential. The
parameters of the Wood--Saxon potential were adjusted to previous data. 
This includes
the bulk of the spectral strength, located in the mean field (MF)
region up to a momentum of $\approx$ 250~MeV/c. 
 However, the position of the SRC correlated peak at large momenta is not
well reproduced by calculations~\cite{Frick04}.
The latter represents only a fraction of the total number of nucleons in the
system but it is the part probed in the experiment under consideration.
Given the above considerations regarding the results of
Refs.~\cite{danielaPRL,Frick04}, it is appropriate to extract
the spectral function in the correlated region from
the experimental data. 
 We choose to employ $S^h_{LDA+WS}(k,E)$ for the MF part
but to substitute it in the SRC region with a parameterization in terms
of a Lorentzian energy distribution
\begin{equation}
 S^h_{corr}(k,E) = 
   { C \;  e^{- \alpha \, k} \over [E - e(k)]^2 + [\Gamma(k)/2]^2} \; ,
 \label{eq:Shcorr}
\end{equation}
which was fitted to the experimental results for ${}^{12}{\rm C}$ in parallel
kinematics.
Linear functions of momentum $e(k)$ and $\Gamma(k)$ were sufficient
to obtain a fit of the region around the top of the correlated ridge,
where the experiment appears to be free from FSI effects. 
The shape assumed for the spectral function at very high energies is then
determined by Eq.(\ref{eq:Shcorr}).
%
 The quality of the fit can be judged
from Fig.~\ref{fig:parall_exp}.
The full spectral function employed in Eq.~(\ref{eq:TotRes}) is 
\begin{equation}
 S^h(k,E) = 
  \left\{
      \begin{array}{lll}
         S^h_{corr}(k,E) &~& \hbox{for $k > $230~MeV/c and $E > $19~MeV,}   \\
         S^h_{LDA+WS}(k,E)  &~& \hbox{otherwise.}
      \end{array}
  \right.
 \label{eq:Shtotal}
\end{equation}
This choice provides a smooth transition between the mean-field and the
correlated parts and gives well behaved distributions of energy and momentum,
obtained by integrating $S^h(k,E)$ over ${\bf k}$ and $E$ respectively.
The total number of protons predicted by Eq.~(\ref{eq:Shtotal}) is 5.97.
For ${}^{12}{\rm C}$ the same spectral function was
employed for both protons and neutrons.
For ${}^{197}{\rm Au}$ an analogous construction of $S^h(k,E)$ was
done, which used the same $S^h_{corr}(k,E)$ of Eq.~(\ref{eq:Shcorr})
normalized according to the correct number of protons or neutrons.

Figure~\ref{fig:parall_exp} compares the experimental data for
parallel kinematics with the results of Eq.~(\ref{eq:TotRes}).
 The experiment agrees with parameterization, Eq.~(\ref{eq:Shcorr}), at
low missing energies. 
 At larger energies the total experimental strength starts increasing
again and becomes more than an order of magnitude larger than
the direct signal. 
The fact that this deviation starts sharply at the pion production
threshold, $E_m \approx$150~MeV, indicates that ($e,e'p\pi$) gives a large
contribution. This reaction produces a distorted image of the correlated 
region at
larger missing energies (due to the extra energy necessary for creating
the pion).
At even larger missing energies, other mesons can be produced and the
experimental $E_m$-distribution becomes rather flat.
As the  missing momentum increases, the regions dominated by correlated 
nucleons and pion production tend to overlap.
We note that as long as multiple rescattering effects can be neglected,
quantum interference is not expected to play a role
since these reaction mechanisms lead to different
particles in their final states. 

\begin{figure}[!t]
\vspace{.15in}
  \begin{center}
    \includegraphics[width=0.6\linewidth]
                                  {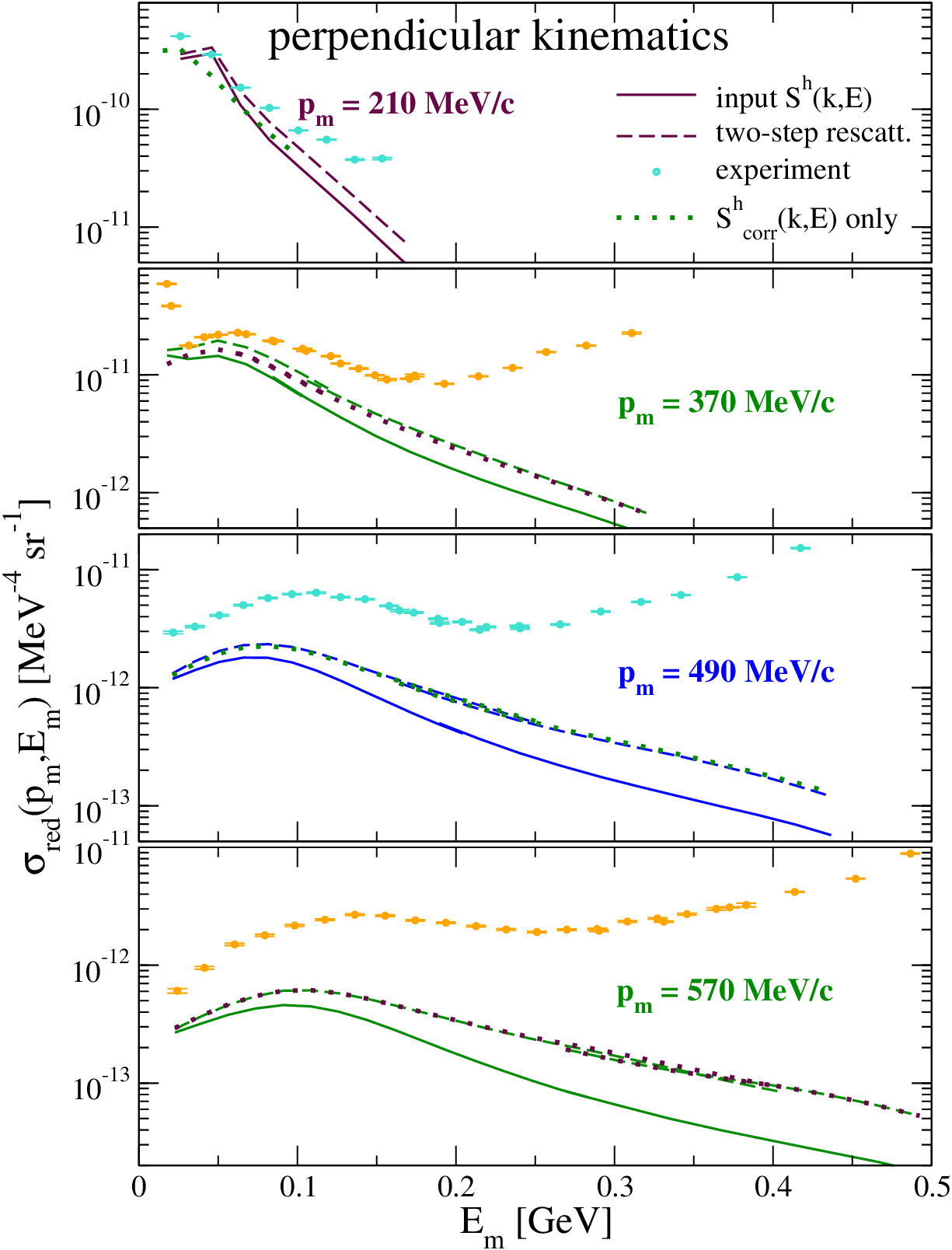}
    \caption{ \label{fig:perp_exp}
       Theoretical results for the reduced cross section of
      ${}^{12}{\rm C}$  obtained in
      perpendicular kinematics (dashed line) compared to the experimental
      results of Ref.~\cite{danielaDiss}.
       The full line shows the model spectral function of
      Eq.~(\ref{eq:Shtotal}) employed in the calculations.
       The dotted lines show the results obtained by neglecting the
       rescattering from the MF region, $S^h_{LDA+WS}(k,E)$, in
       Eq.~(\ref{eq:Shtotal}).
        This becomes indistinguishable from the full calculation
       for $p_m >$~450~MeV/c.
}
\end{center}
\end{figure}

The experimental data are found to be sensibly 
larger in perpendicular kinematics than in parallel ones.
 The discrepancy between the two cases increases with
the missing momentum and reaches one order of magnitude 
at very large $p_m$'s.
Due to the much larger redistribution of spectral strength the valley
between the SRC and meson production regions is also filled and 
it is no longer possible to distinguish them.
 The data for perpendicular kinematics are compared to 
the results of Eq.~(\ref{eq:TotRes}) in Fig.~\ref{fig:perp_exp}.
In this case, the rescattered strength is large and
affects the reduced spectral function already at the top of the
SRC tail.
%
%
 Kinematically, this can be understood by observing that for perpendicular
kinematics a nucleon can loose energy in a rescattering event and still
be detected with a missing momentum larger than its initial momentum.
This results in moving strength from regions where the spectral function is
large to regions where it is small, hereby submerging the direct signal.
The shift is large enough that measurements in the correlated region can
be affected by events originating from the MF orbits~\cite{BaLap04}.
  This is displayed by the dotted lines in Fig.~\ref{fig:perp_exp}, which
have been obtained by setting $S^h_{LDA+WS}(k,E)=0$ in Eq.~(\ref{eq:Shtotal}).
The effect of neglecting rescattering from the MF strength is appreciable for
missing momenta lower than 500~MeV/c, but becomes negligible above it.
For parallel kinematics a similar comparison shows
no visible effects indicating that rescattering moves strength
{\em only} within the correlated region itself.
Fig.~\ref{fig:perp_exp} shows that the Eq.~(\ref{eq:TotRes}) can account
for the differences between parallel and perpendicular kinematics
for $E_m \approx$~50~MeV and $p_m <$~350MeV/c.
However, it strongly under-predicts the experiment over the larger
part of the correlated region.
  For $p_m=$600~MeV/c, the cross section is predicted to be 50\% 
larger than the direct process, whereas the experiment is
about of one order of magnitude above.
Moreover, the experimental data for perpendicular kinematics show a further
rise for $E_m \ge$~200~MeV, in the region of meson production.
We conclude that two-step rescattering represents only a fraction of the
total FSI present in perpendicular kinematics.
A possible contribution that could bring theory and data closer together
would be that of MEC currents that involve the excitation of a $\Delta$
or higher resonances; these are known to be sensitive to transverse degrees
of freedom.
 Besides, the effects of multiple rescattering and IC can become relevant,
especially for heavy nuclei~\cite{NicolaevGL,Ciofi4He}

In conclusion, we have carried out a comparison of the experimental
($e,e'p$) data  at large $p_m$-$E_m$
with first theoretical predictions of rescattering effects for the same
kinematics.
 The results are understood in terms of kinematical constraints and
confirm the expectation that, for light nuclei and properly
chosen parallel kinematics, the effects of FSI can be small
for missing energies up to the $\pi$ emission threshold.
 In perpendicular kinematics the agreement between data and theory
is less favorable. This suggest that additional ingredients, 
of transverse character, such as MEC, should be included in the calculations.

\ack
This work is supported by the Natural
Sciences and Engineering Research Council of Canada (NSERC),
by the Schweizerische Nationalfonds (NSF)
and
by the ``Stichting voor Fundamenteel Onderzoek der Materie (FOM)'',
which is financially supported by the ``Nederlandse Organisatie voor
Wetenschappelijk Onderzoek (NWO)''.



\end{document}